\documentclass[11pt]{article}
\usepackage{moriond,epsfig}

\def\be{\begin{equation}}
\def\ee{\end{equation}}
\def\bea{\begin{eqnarray}}
\def\eea{\end{eqnarray}}

\def\url#1{{\ttfamily\def\/{/\discretionary{}{}{}}#1}}

\def\Msun{\mbox{M$_\odot$}}

\def\mathnew{\mathsurround=0pt}
\def\simov#1#2{\lower .5pt\vbox{\baselineskip0pt
    \lineskip-.5pt\ialign{$\mathnew#1\hfil##\hfil$\crcr#2\crcr\sim\crcr}}}

\def\'#1{\ifx#1i{\accent"13\i}\else{\accent"13#1}\fi}

\begin{document}
\vspace*{4cm}
\title{DARK SATELLITES AND COSMIC REIONIZATION}

\author{ Andrea V. Macci\`o,
Ben Moore, Joachim Stadel \& Doug Potter}

\address{Institute for Theoretical Physics, University of Z\"urich }

\maketitle\abstracts{
A possible explanation of the present discrepancy between the abundance 
of galactic subhaloes predicted by N-Body simulations with those observed
in the Local Group is presented. 
We study the impact of an early reionization on the baryonic 
component of the Universe using SPH simulations on group and galactic scales.
We use a simplified model for reionization described as an instantaneous increment of the IGM 
temperature (with $10^5<T_{IGM}<10^6$ K) at a given redshift ($z_r
\geq 8$).
We find that a key role is played by compton cooling 
(interaction between hot electrons and the CMB photons); at high redshift 
($z_r>10$) this cooling is very efficient and it is able to counteract any heating
of the gas within few Myrs ($\approx 70$). This means that a late
reionization is needed ($z_r<9$) to sufficiently reduce the number of
luminous dwarf satellites around our Galaxy.
For a reionization redshift $z_r=8$ and a reionization temperature of
$T_{IGM} \approx 10^5$ K we are able to reproduce the observed number of
Local Group dwarf galaxies in our simulations.}

\section{Introduction}

Cold Dark Matter (CDM) models are the current paradigm for the formation of
Large Scale Structure and they successfully explain most properties of the local and
high-redshift Universe. 
One of the (few) problems of such models is their failure to reproduce the 
abundance of satellites with a circular velocity of $V_{c} \approx 10-30$  
km sec$^{-1}$ observed within the Local Group.
High-resolution N-body simulations reveal a large number of embedded subhaloes 
that survive the collapse and virialization of the parent structure 
(Klypin et al 1999; Moore et al 1999). 
Such large numbers of subhaloes are in apparent discrepancy
with the number of faint satellites observed in the Local Group by an order of 
magnitude (i.e. Macci\`o et al 2006a).
Many authors have pointed out that accretion of gas onto low-mass
haloes and subsequent star formation are inefficient in the presence of
a strong  photoionizing background (i.e. Quinn, Katz, \&  Efstathiou 1996). 
The ultraviolet radiation and mechanical energy that preheated and reionized most 
of the hydrogen in the intergalactic medium (IGM), ending the ``dark ages'', was 
probably the result of an early generation ($z = 10-15$) of subgalactic stellar 
systems aided by a population of accreting black holes (Madau \& Rees 2001). 
This picture, consistent with the hierarchical nature of structure formation in 
the Universe, is supported by the recent analysis of the third year's data from 
the WMAP satellite suggesting that the Universe was reionized at redshift 
$z_{ion} \simeq 10$ (Spergel et al 2006).
Reionization raises the entropy of the gas, preventing it from accreting onto 
small dark matter haloes and lengthening the cooling time of that gas which has 
accreted (Bullock et al (2000) see also Kravtsov, Gnedin \& Klypin 2004).
Moreover, Benson \& Madau (2003; BM03 hereafter) suggested
that winds from pregalactic starbursts and
miniquasars may pollute the intergalactic medium (IGM) with metals and raise
its temperature to a much higher level than expected from photoionization alone and
so further inhibit the formation of early galaxies.

The detailed history of the Universe during and soon after these crucial
formative stages depends on the power spectrum of the density fluctuations on
small scales and on a complex network of poorly understood ``feedback'' mechanisms.
For the first time, in order to elucidate the effect of all the cooling mechanisms 
that can act at high-redshift against the photo evaporation of gas in DM haloes 
we use several ``toy'' models for reionization within high-resolution hydrodynamic 
N-body simulations of structure formation.
We particularly focus our attention on the Compton cooling (due to scattering
between hot electrons and CMB photons, CC hereafter) which we argue 
is of critical importance for forming structures before and during reionization.

\section{Simulations}
\label{subsec:prod}

The simulations were performed with GASOLINE, a multi-stepping, parallel 
TreeSPH $N$-body code (Wadsley et al. 2004).
We include radiative and Compton cooling for a primordial 
mixture of hydrogen and helium. The star formation algorithm is based on 
a Jeans instability criteria. The code also includes supernova
feedback as described by Katz (1992), and a UV background following Haardt \&
Madau (1996) (see Macci\`o et al 2006a and 2006b for more details).

We run two different kinds of simulations: in the first set of simulations
we selected a candidate halo from an lower resolution, dark matter
only, $\Lambda$CDM simualtion ($\Lambda$=0.7, $\Omega_0$=0.3, $\sigma_8$=0.9)
and re-simulate it at higher resolution using the volume 
renormalization technique (Katz \& White 1993) including a gaseous component
within the entire high resolution region.
This halo is a group-like object ($M \approx 10^{13} \Msun$) at $z=2$ where we 
stop simulating. In these simulations the mass of the dark matter and gas particles is
$4.7 \times 10^7$M $_{\odot}$ and $1.1 \times 10^7$M$_{\odot}$ respectively
and we have more than 223000 dark and 97000 gas particles in the high resolution region.
These group simulations were used to explore the parameter space of our reionization
scheme to find the most intriguing models for the second set of simulations.
The second set of simulations are on the galactic scale. 
We selected a candidate galactic mass halo ($M \approx
10^{12} \Msun$) and resimulated it at higher resolution and including a gaseous component
within the entire high resolution region. The mass per particle of the dark matter
and gaseous particles are respectively
$m_{d} = 1.66 \times 10^6 M_{\odot}$ and $m_g = 3.28 \times 10^5 M_{\odot}$.  
The SPH simulations with cooling are computationally expensive and we are 
forced to stop the full calculation once the galaxy has formed, at a redshift 
$z=1.5$. (The parallel calculation is dominated by the few remaining gas particles which
need extremely small time steps in order to satisfy the Courant criterion). 
In order to study the dynamical evolution of the stellar 
and dark matter satellites we have evolved the simulation to the present epoch
without following the remaining gaseous particles - we turn them into collisionless
particles and treat only their gravitational interactions. We do not believe that
this influences any of our conclusions since (i) most of the gas has already  turned into
stars between a redshift z=5.5 and z=2.5, and (ii) continuing to
include cooling of the remaining gas would only allow us to resolve higher density gas clouds.
For these galaxy simulations we have 7 runs: one without reionization,
five with reionization for different $z_r$ and $T_r$ values and
finally one pure dark matter run.
A complete list of simulation parameters is contained in table 1.
To find bound structures in our simulations we have used the public
available halo-finder SKID (Stadel 2001, see Macci\`o et al. (2006) for a 
discussion on results resolution dependence and for a more detailed 
description of SKID parameters).
%
%
\begin{table}[t]
\caption{Parameter list of simulations. We also ran an high
  resolution pure dark matter run ($G_0$) with a mass resolution of $M_d=5.8
  \times 10^5 \Msun$.}
\vspace{0.4cm}
\begin{center}
\begin{tabular}{|l|l|l|l|l|l|} 
\hline
{Name} & {Reion} & {$z_r$} &  {$T_r$ (K)}  & {$M_{d} /\Msun $} &
{$M_{g}/\Msun$} \\

\hline

G$_1$ & No  & -- &  -- &$1.6\times 10^6$ &  $3.3\times 10^5$  \\

G$_2$ & Yes & 8  &  $1\times 10^5$    & $1.6\times 10^6$ &  $3.3\times 10^5$  \\

G$_3$ & Yes & 8 &   $5\times 10^5$    & $1.6\times 10^6$ &  $3.3\times 10^5$  \\

G$_4$ & Yes & 8 &   $2\times 10^6$    & $1.6\times 10^6$ &  $3.3\times 10^5$  \\

G$_5$ & Yes & 10 &   $5\times 10^5$   & $1.6\times 10^6$ &  $3.3\times 10^5$  \\

G$_6$ & Yes & 13.5 & $5\times 10^5$ & $1.6\times 10^6$ &  $3.3\times 10^5$  \\

\hline
\end{tabular} 
\end{center}
\end{table} 

\subsection{Compton Cooling and reionization modeling}

\begin{figure}
\begin{minipage}{0.30\textwidth}
\begin{center}
\includegraphics[width=\textwidth]{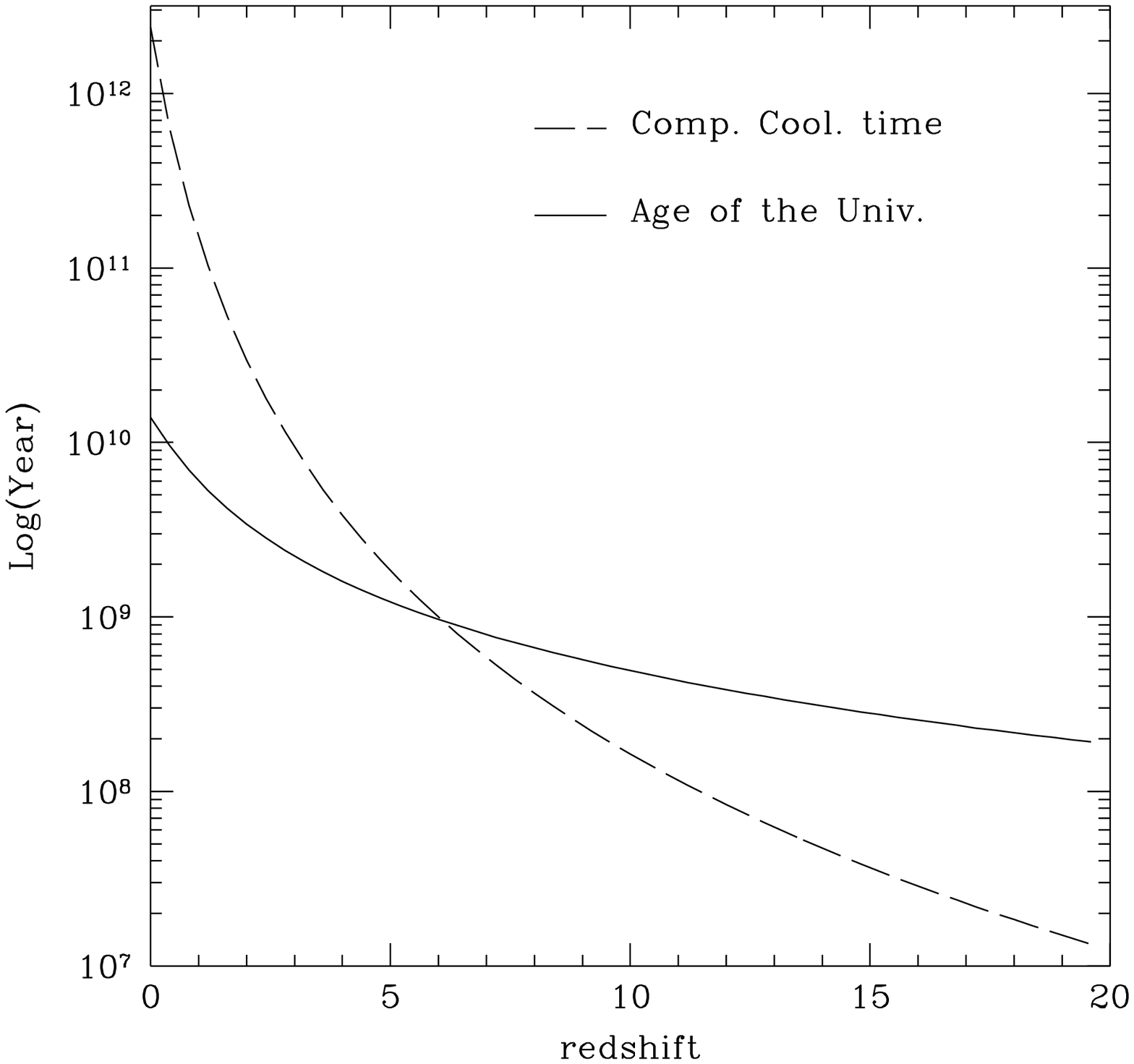}
\end{center}
\end{minipage}
\hfill
\begin{minipage}{0.32\textwidth}
\begin{center}
\includegraphics[width=\textwidth]{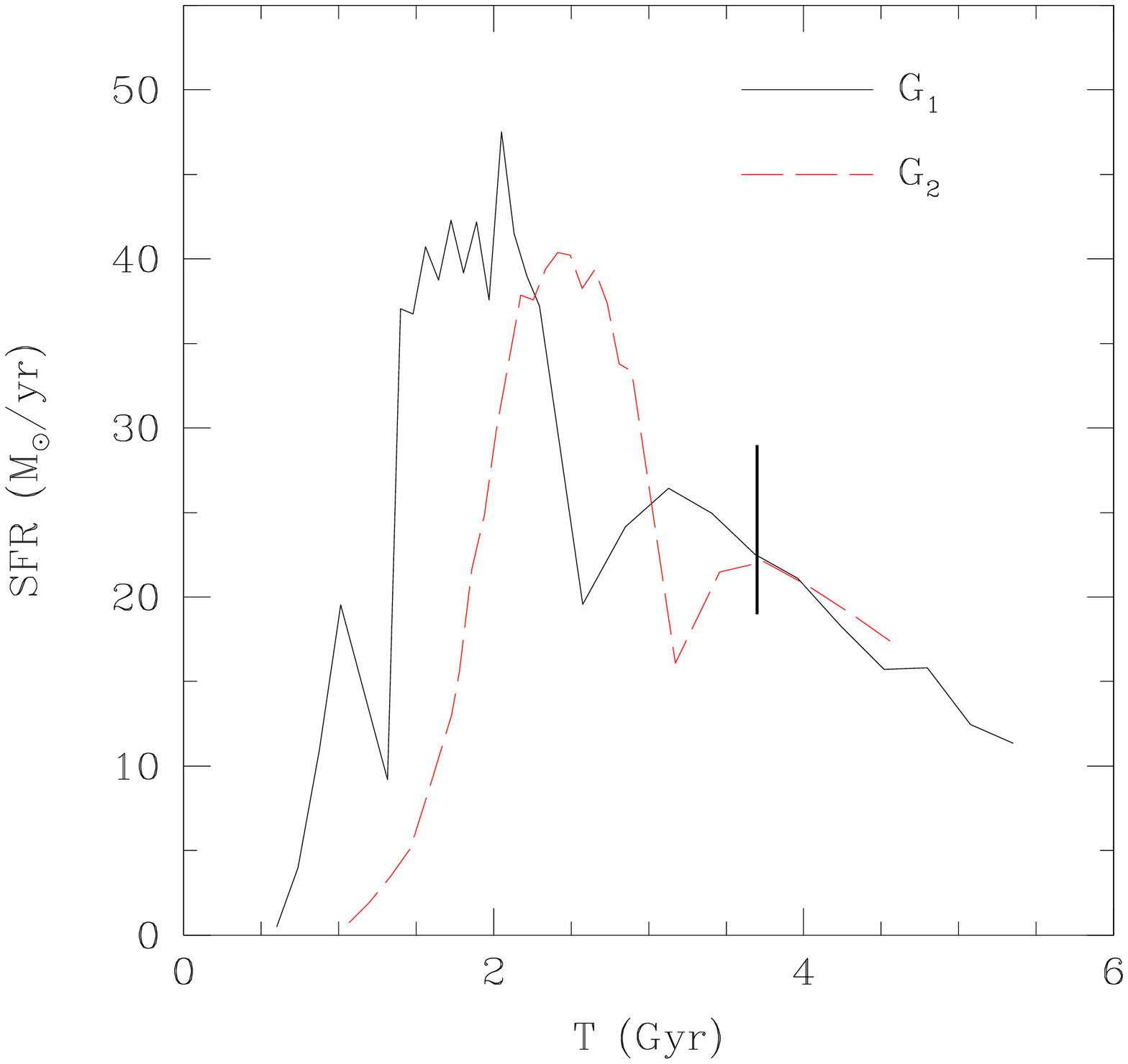}
\end{center}
\end{minipage}
\hfill
\begin{minipage}{0.30\textwidth}
\begin{center}
\includegraphics[width=\textwidth]{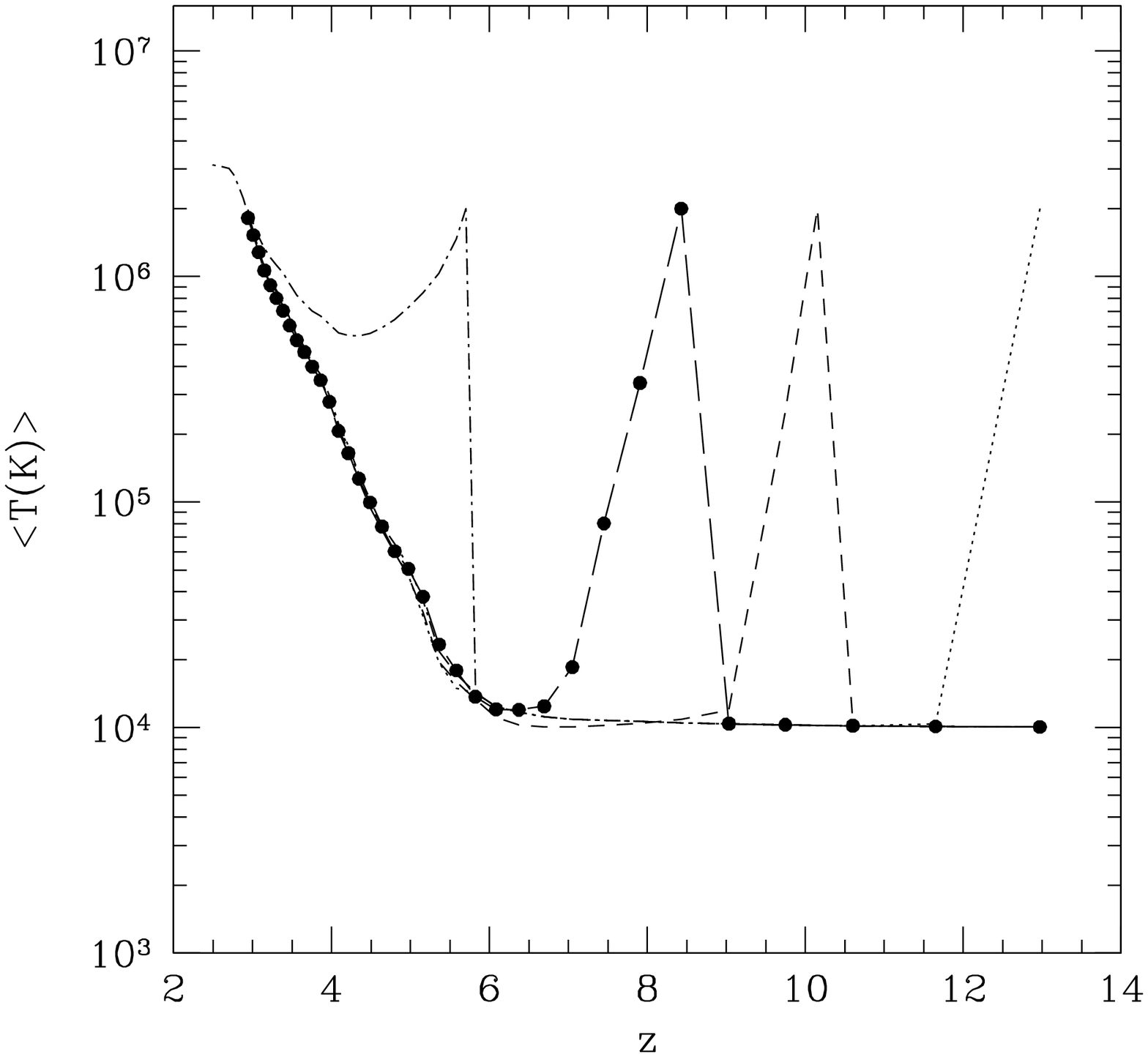}
\end{center}
\end{minipage}
\caption{Left: Comparison between the CC time (dashed line) and the
  age of the Universe (solid). Middle: Star Formation Rate in
  different models. Right: Evolution of the mean temperature of gas particles in run with
different reionization epochs. Solid line is for the no reionization run.}
\label{fig:one}
\end{figure}

When photons of low energy $h\nu$ pass through a thermal gas of
non relativistic electrons ($h\nu \ll kT_e \ll m_ec^2$) they scatter
with the Thompson cross section.
Provided that the gas (electrons) temperature is greater than the
photon's temperature ($T_e \gg T_{\gamma} =2.7(1+z)K$), the gas will cool against
the microwave background on the timescale $t_{comp}(z) = 3kT_e / <dE_e/dt>$.
The evolution of this cooling time with redshift can be easily computed (White 1996):
\begin{equation}
t_{comp}(z) = {{2.40 \times 10^{12}} \over {(1+z)^4}} ~~ yrs.
\end{equation}
Note that this timescale is {\it independent} of the temperature and density of the gas.
A comparison between the behavior of $t_{comp}$ and the age of the Universe
with redshift is shown in figure \ref{fig:one} left panel. 
It is clear that for $z>8$ this cooling time is very short and can not
be neglected, because it can counterbalance any heating of the gas particles.

On the other hand, strong feedback is necessary to prevent too many baryons from turning into
stars as soon as the first levels of the clustering hierarchy collapse.
The required reduction of the stellar birthrate in haloes with low circular
velocities may naturally result from the heating and expulsion of material due
to quasar winds and repeated SN explosions from an early burst of star
formation. Furthermore as blast waves produced by miniquasars and protogalaxies
propagate into intergalactic space, they may drive the IGM  to much
higher temperature than expected from the photoionization background
responsible for its reionization (i.e. Cen \& Brian 2001, BM03). 
The latter authors have also computed the thermal evolution of the IGM when it
is rapidly preheated at a given redshift. Observations of the temperature of
the IGM at $z \simeq 3$ allowed them to rule out models in which this
preheating occurs later than some epoch ($z \approx 8$)  or sets the IGM
higher than a certain temperature (million degrees).

Starting from their findings we have modeled reionization as an instantaneous 
increase of the IGM temperature (van den Bosch et al 2003). 
We have two free parameters: the reionization
redshift ($z_r$) and the reionization temperature ($T_r$). Even if this 
kind of reionization model is quite extreme, as it implies a big 
star formation burst and then a passive evolution of the IGM, it can
provide useful hints on the effects of reionization on dwarf galaxies.
Using the group-like run, we have explored a wide range for these parameters. 
One of our results is that in this scenario, a key role is played by
the CC. 
This can be seen in figure ~\ref{fig:one} (right panel), where
results for the Group-like simulations are shown.
We calculated the average temperature of all the 
gas particles for each $(T_r;z_r)$ run. From a comparison of the $<T>$ in runs with different 
reionization epochs (keeping  $T_r$ fixed to $2 \times 10^6$ K) we can gauge
the efficiency of CC with redshift. For a reionization models with $z_{r} > 9$ 
the gas temperature drops quickly from $2\times 10^6$ K to 
$10^4$ K (below this temperature electrons and protons start to
recombine, effectively switching off CC) and then follows the same
track as the run without reionization. 
The very short cooling time at these redshifts reduces the heating effect of 
reionization on the subsequent state of the the gas.
The situation is different for $z_r=5.6 $, in this case $t_{comp}>t_H$ and so the 
temperature of the gas particles remains well above $10^6~K$.
For $z_r=8$ we have an intermediate case where the temperature decreases from 
$10^6~K$ to $10^4~K$ but not as quickly as in the $z_r>9$ runs.

\section{Numerical Results and Comparison with Observations}

Starting from the group simulation findings we ran 6 different
reionization models of the same galaxy: one with no reionization and 5
with various combination of $(z_r:T_r)$ (see Table 1).
The first effect of reionization can be seen in the evolution of the Star Formation Rate (SFR)
in different runs. For example, in the $G_2$ model the peak of the
SFR is delayed with respect to the standard run ($G_1$) by about one
Gyr as shown by figure \ref{fig:one} (middle panel).

We evolved all runs with hydro until z=1.5. At this redshift we 
marked all of the star particles inside our simulations, so we were able
to follow their evolution in the pure gravitational run from $z=1.5$ to $z=0$.
In order to quantify the effects of reionization we need to compute
how many luminous satellites survive up to the present epoch.
We decided to use the circular velocity ($V_{c}$) function (CVF) of
luminous haloes (i.e. haloes containing stars) to compare the different runs.
$V_{c}$ is less affected by the lack of hydro evolution because its
value is set by the overall satellites mass distribution which is dark matter
dominated on large scales ($>10$ kpc).
Using circular velocity also makes possible a more direct comparison with
observed satellites abundance around the Local Group.
For this comparison we gathered together data for the Local Group taken from
Mateo et al (1998), Odenkirchen et al 2001 and Kleyna et al (2005); we
used the results of Kazantzidis et al (2005) to convert
three-dimensional velocity dispersion into circular velocity.
We considered the number of satellites per central
galaxy instead of considering the local group as a whole. This makes the
comparison with simulations straightforward. In this sample there was not
attempt to decide whether a satellite is bound to its central object.
Satellites were counted if they lay within a certain radius from
the center of their parent galaxy. We arbitrarily chose a radius $r=
280$ kpc for the counts.
In all the figures, the observational data set is shown as black
triangles with error bars (one standard deviation).

\begin{figure}
\begin{minipage}{0.30\textwidth}
\begin{center}
\includegraphics[width=\textwidth]{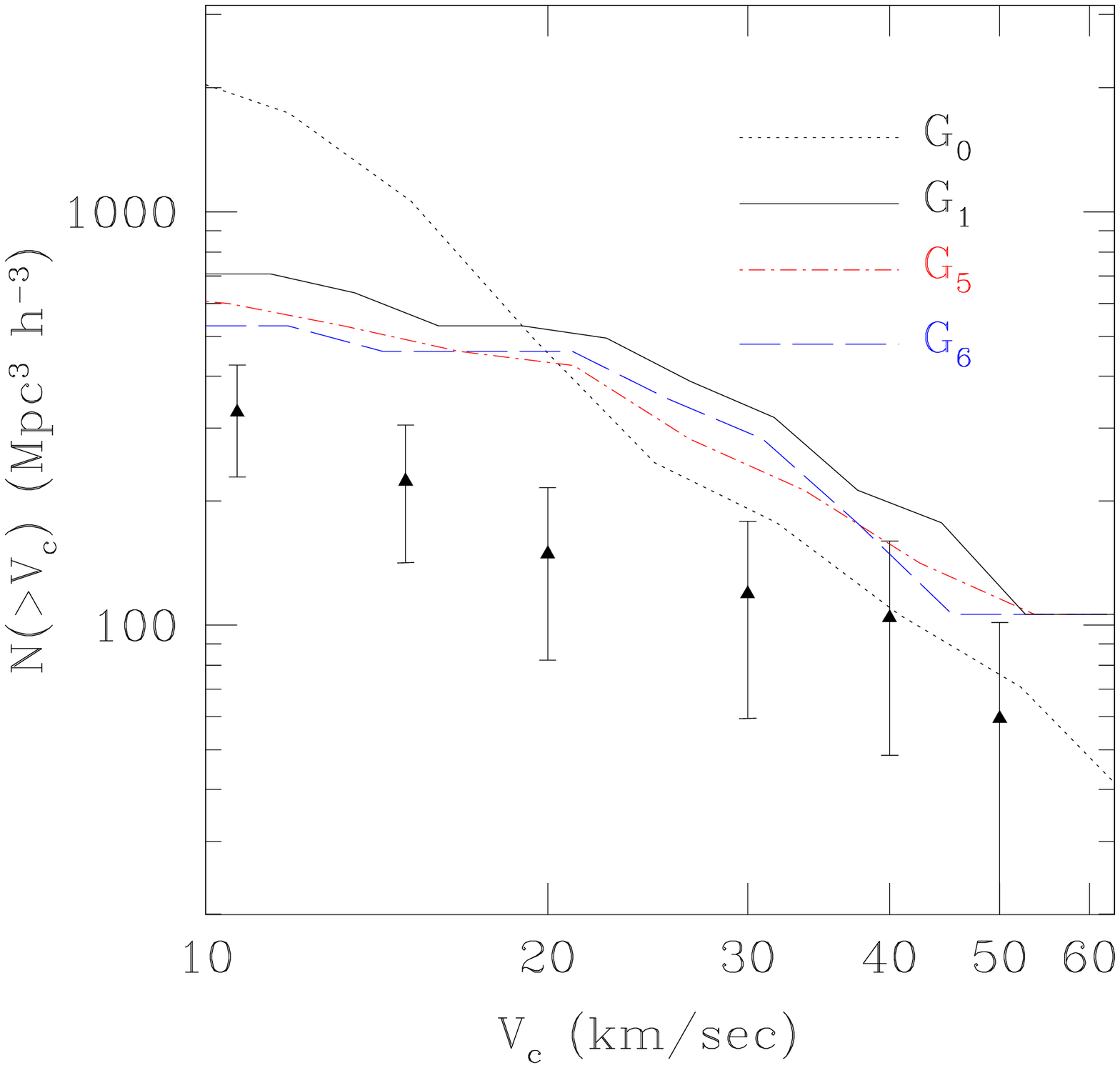}
\end{center}
\end{minipage}
\hfill
\begin{minipage}{0.30\textwidth}
\begin{center}
\includegraphics[width=\textwidth]{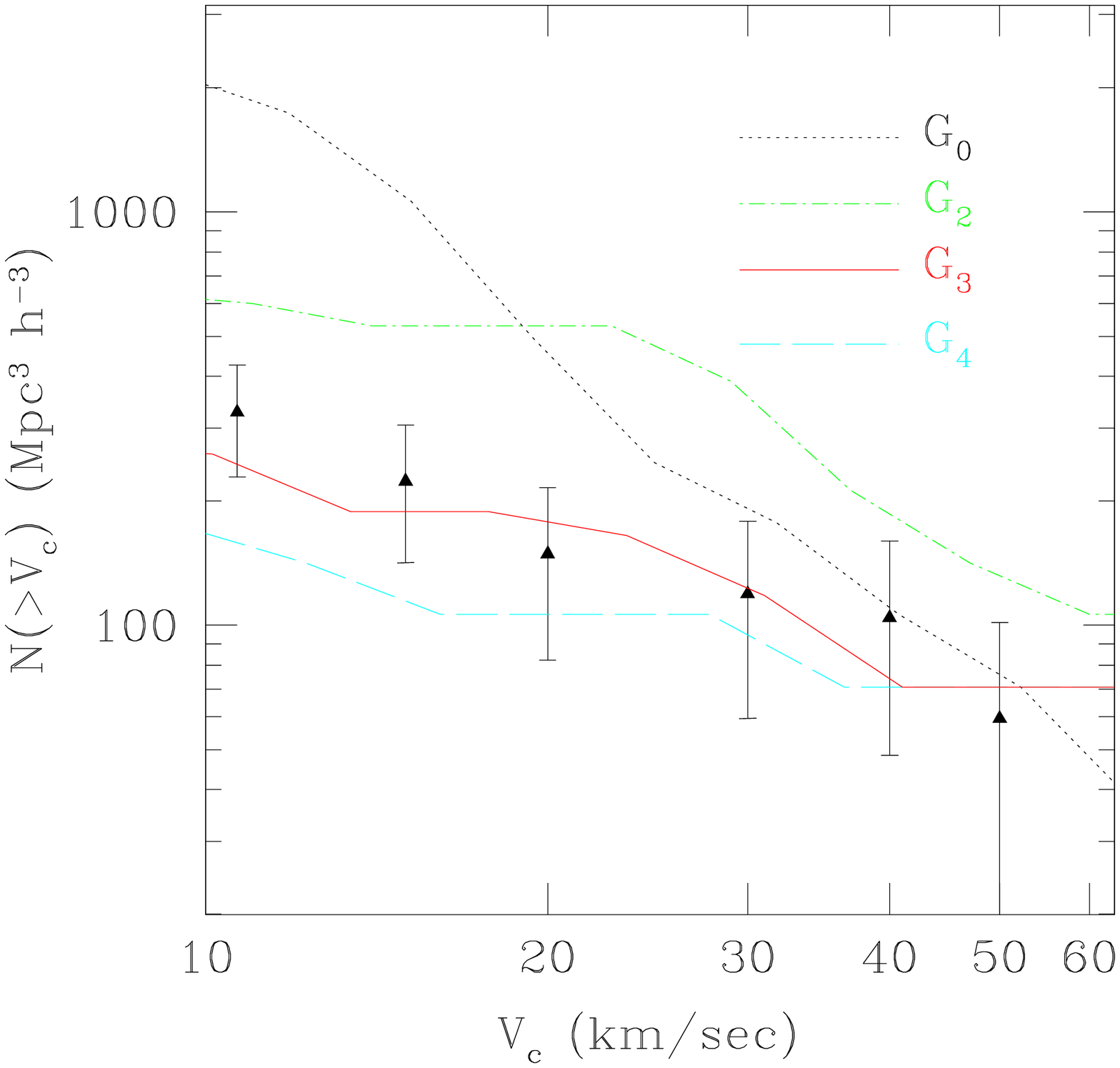}
\end{center}
\end{minipage}
\hfill
\begin{minipage}{0.30\textwidth}
\begin{center}
\includegraphics[width=\textwidth]{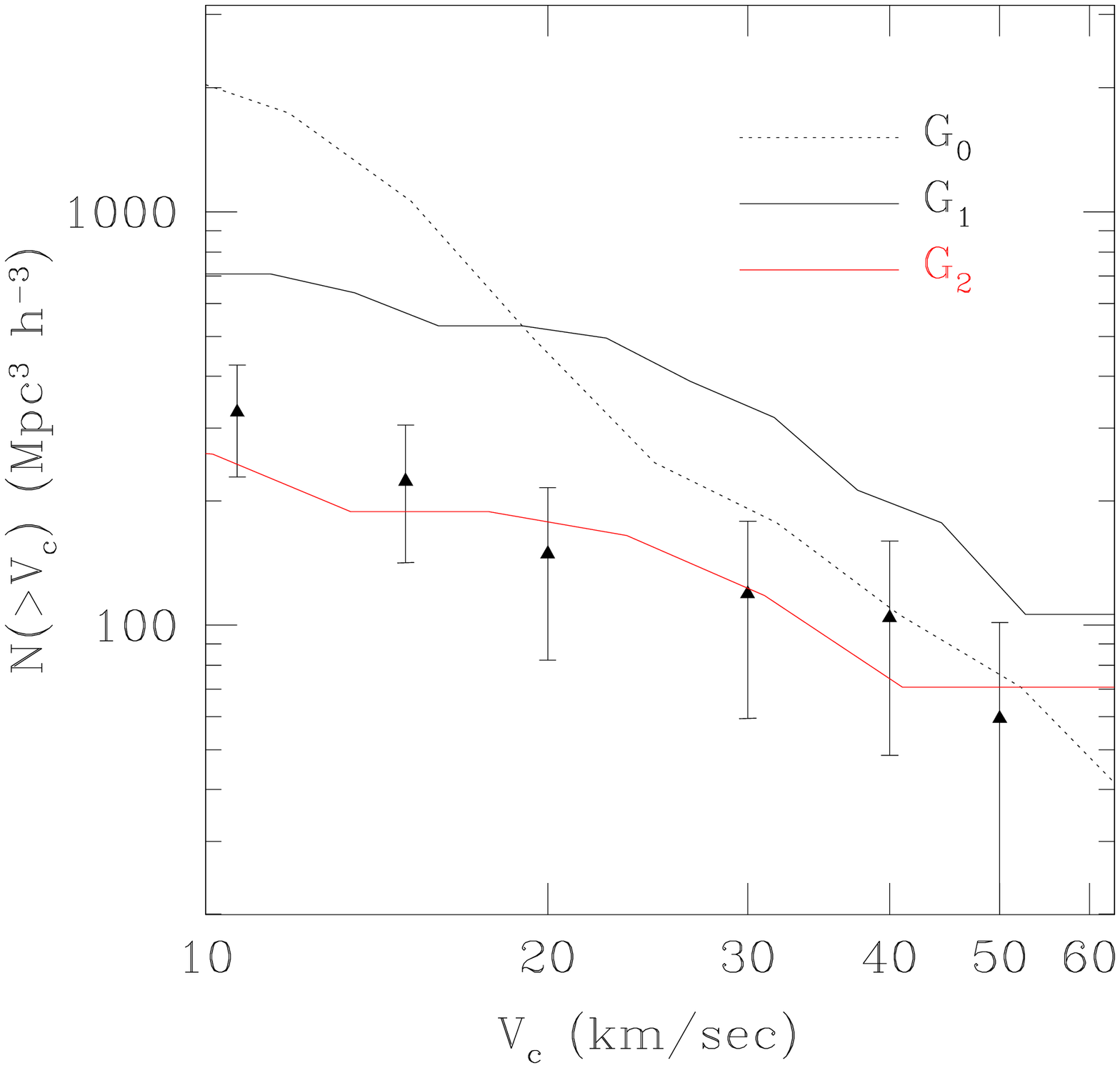}
\end{center}
\end{minipage}
\caption{Cumulative circular velocity distribution function of luminous 
  satellites within 280 kpc from the main halo center for different reionization models.
  The black dotted line represents results for dark matter halos in a
  pure Nbody simulation.  Black triangles with error bars show average
  results for Milky Way and Andromeda satellites (see text for details).}
\label{fig:two}
\end{figure}

The left  panel of figure \ref{fig:two} show the CVF for
luminous satellites for the model without reionization ($G_1$, solid
black line) and for the two high redshift reionization models ($G_5$
and $G_6$, blue and red lines). The dotted line shows results for a
pure dark matter run ($G_0$). As noticed in Macci\`o et al
2006 the main effect of hydro is to enhance 
the survival probability of satellites due to their steeper density
profile (resulting from the presence of baryons).

\begin{figure}
\begin{minipage}{0.45\textwidth}
\begin{center}
\includegraphics[width=0.6\textwidth]{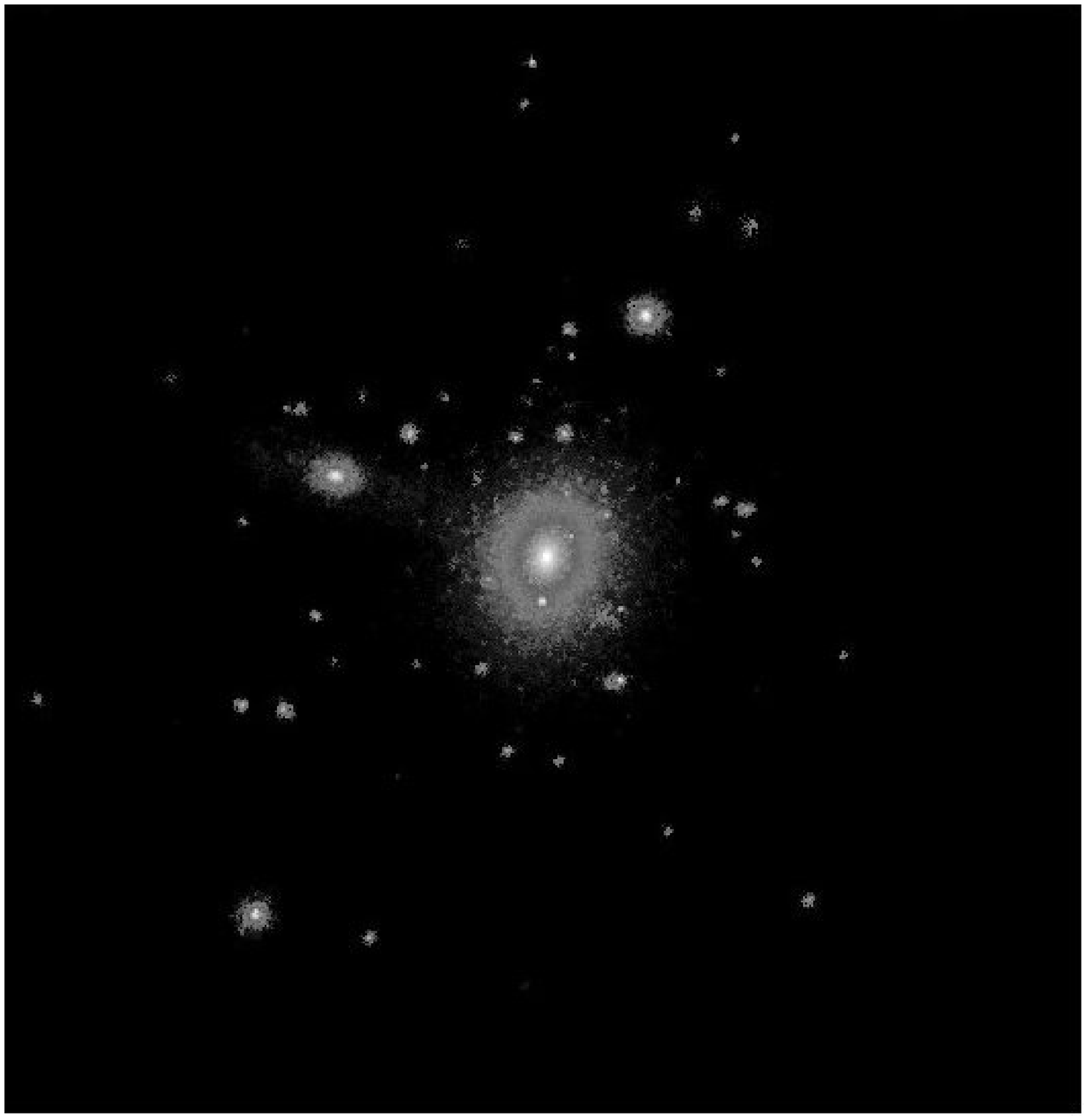}
\end{center}
\end{minipage}
\hfill
\begin{minipage}{0.45\textwidth}
\begin{center}
\includegraphics[width=0.6\textwidth]{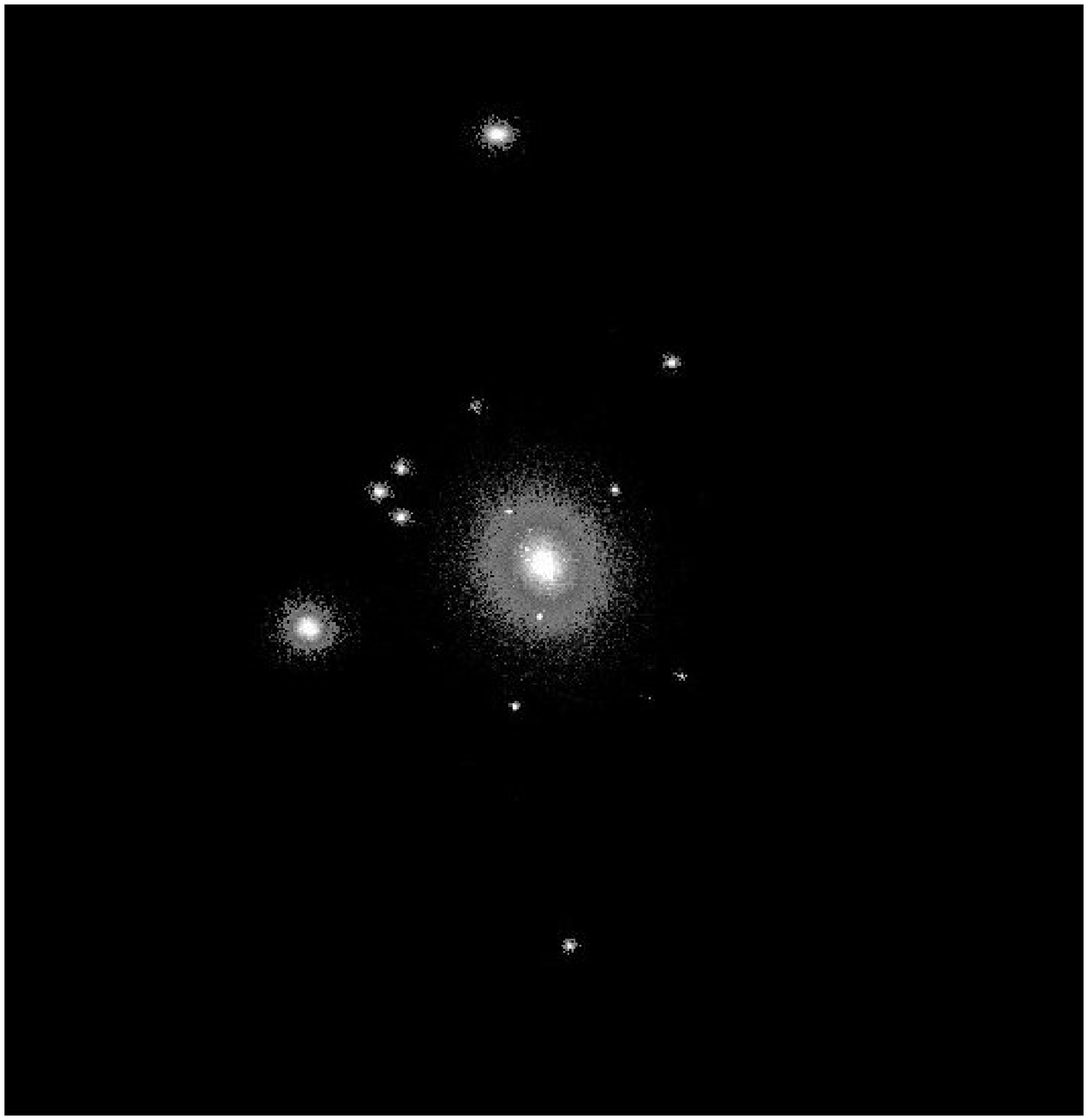}
\end{center}
\end{minipage}
\caption{Left: $G_0$ run dark matter satellites overdensity at
  z=0. Right: $G_3$ stellar satellites overdensity at z=0}
\label{fig:three}
\end{figure}

On the other hand as already discussed in the previous section, it is 
clearly visible that models with high redshift reionization ($z_r \ge
10$) are unable to deplete the formation of visible satellites, mostly
for the effect of the compton cooling.
In figure \ref{fig:two} (central panel) are shown the results for low-$z$ reionization
models. In this case modifications to the CVF depend on the
virialization temperature. For $T_r=10^5$ K ($G_2$) the CVF is very similar
to the one without reionization. For $T_r=2 \times 10^6$ K ($G_4$) the
effect of reionization is too strong. At $z=0$ only 6 satellites
survive and 4 of them have $V_{c}>$30 km/sec. 
Model $G_3$ is the one showing the better agreement with
observations. In this case the gas accretion depletion is enough to reconcile the
observed and simulated number of satellites at $z=0$.
The difference in the star satellites number and distribution can be
also appreciated by looking directly at the density map of stars. In
the left and right panels of figure \ref{fig:three} we present the stellar
satellites overdensity in $G_0$ and $G_3$ models respectively.

The main result of this work is summarized in the right panel of figure \ref{fig:two}
where a comparison between model $G_0$, $G_1$ and $G_3$ is presented.
This picture clearly shows that the $G_3$ choice of reionization
temperature and redshift are able to correctly reproduce 
both the shape and the normalization of the CVF ofstar haloes as observed in our neighbors.

\section{Conclusions}

We have tested the impact of a reionization on the abundance of visible haloes
around a Milky Way like simulated galaxy. We have used high resolution
hydrodynamical simulation in which we implemented a simple treatment
of reionization.
In our model, cosmic reionization  is parametrized as an instantaneous 
increase of the temperature of the IGM. Two free parameters are in the
model: $T_r$ the temperature of the IGM after reionization and $z_r$
the redshift at which it occurs.
Using a first set of simulations of a group like object ($M \approx 10^{13} \Msun$) 
we inspected a wide range of parameters for our model. 
Our first finding is that a key role is playing by the Compton
Cooling (between hot electrons of the IGM and CMB photons). If the
reionization occurs at high redshift ($z>9$), the compton cooling time is very
short with respect to the hubble time (cfr fig \ref{fig:one} left
panel). Thus CC is able to cool down the IGM in few Myrs ($\approx 100$), counteracting
any reionization effect. To avoid this high cooling we need a later
reionization ($z_r \approx 9-8$) as suggested by WMAP third year
results in contrast with WMAP first year results ($z_r \approx 17$)
(Spergel et al 2006).
Starting from these results we ran 6 high resolution ($N_{gas}
\approx 6 \times 10^5$) hydro galaxy simulations spanning the parameter
space ($z_r:T_r$). 
We showed that with later reionization ($z_r=8$) and a mildly high IGM
temperature $T_r=5 \times 10^5$ K (in agreement with observational data
on quasar absorption lines around $z=3$, BM03) 
the total number of luminous satellites surviving at the present epoch
is in agreement with the dozen of dwarf galaxies observed around the
Milky Way. In this way it is possible to reconcile the  flat
luminosity function observed within the Local Group
with the steep mass function of haloes and subhaloes within the cold dark matter model.

\section*{Acknowledgments}
The authors acknowledge J. Wadsley for development of the GASOLINE code and 
thank him for its use in this work. 
All the numerical simulations were performed on the zBox2 supercomputer
(http://www-theorie.physik.unizh.ch/$\sim$dpotter/zbox2/) at the University of Z\"urich.

\end{document}